\documentclass[12pt,epsf]{article}
\newlength{\abstractwidth}
\usepackage[dvips]{graphicx}
\usepackage{subfigure}

\renewcommand{\thefootnote}{\fnsymbol{footnote}}
\renewcommand{\thanks}[1]{\footnote{#1}} 
\newcommand{\starttext}{
\setcounter{footnote}{0}
\renewcommand{\thefootnote}{\arabic{footnote}}}
\renewcommand{\theequation}{\thesection.\arabic{equation}}
\newcommand{\be}{\begin{equation}}
\newcommand{\bea}{\begin{eqnarray}}
\newcommand{\eea}{\end{eqnarray}}
\newcommand{\beq}{\begin{equation}}
\newcommand{\ee}{\end{equation}}
\newcommand{\eeq}{\end{equation}}

\newcommand{\<}{\langle}

\renewcommand{\>}{\rangle}
\def\ba{\begin{eqnarray}}
\def\ea{\end{eqnarray}}

\def\14{{1\over4}}
\def\12{{1 \over 2}}

\def\h3{h^{3\over 2}}

\def\>{\rangle}
\def\<{\langle}

\def\0cc{$\Lambda = 0$}

\begin{document}
\renewcommand{\theequation}{\thesection.\arabic{equation}}
\begin{titlepage}
\bigskip
\rightline{SU-ITP } \rightline{hep-th/}

\bigskip\bigskip\bigskip\bigskip

\centerline{\Large \bf {Wormholes and Time Travel? Not Likely }}

\bigskip\bigskip
\bigskip\bigskip

\centerline{\it L. Susskind  }
\medskip
\centerline{Department of Physics} \centerline{Stanford
University} \centerline{Stanford, CA 94305-4060}
\medskip
\medskip

\bigskip\bigskip
\begin{abstract}
Wormholes have been advanced as both a method for circumventing
the limitations of the speed of light as well as a means for
building a time machine (to travel to the past). Thus it is argued
that General Relativity may allow both of these possibilities. In
this note I argue that traversable wormholes connecting otherwise causally disconnected
regions, violate two of the most fundamental principles physics,
namely local energy conservation and the energy-time uncertainty principle.
\end{abstract}

\end{titlepage}
\starttext \baselineskip=18pt \setcounter{footnote}{0}


 \setcounter{equation}{0}
\section{Wormholes}
A wormhole is a multiply connected spatial geometry that consists
of two ``mouth holes" separated by some distance. The lips of the
mouths are identified so that a shortcut exists through the
wormhole. Going the long way, the distance between the mouths,
call it  $L$,  may be huge or even infinite.  By contrast the
distance $l$, through the wormhole, may be very small.
 According to some authors, such wormholes might be left
over from the big bang, or it might even be possible to construct
them.  They are very
 common, being frequently seen in the press, in science fiction stories, and in physics books for a popular
audience.

Such wormholes, if they exist, would be very interesting windows on distant
regions of space--even regions beyond our cosmic horizon. We might
pass through one, take a look around, and come back to report what
we saw \cite{kip}\cite{greene}. If we were not bold enough to do that, we could send a TV
camera connected to a cable. Wormholes have even been advocated as
a means to escape a dying bubble-universe to a younger safer one \cite{kaku}.
If they existed they could give operational meaning to the
``Multiverse" idea. Finally, they are purported components of time machines \cite{deser}.

In analyzing these claims, it is important to realize that
traversable wormholes (TW's) are not possible in classical physics.
Classical GR does not allow a wormhole such as an Einstein Rosen
Bridge, to stay open long enough for an object to cross from one mouth to the
other. The idea is that quantum mechanics, usually in the form of
the Casimir effect, can allow violations of the positive energy
conditions that prohibit TW's. I am going to give wormholes the benefit of the doubt
and assume that the quantum effects can stabilize them. However, because they do not
exist in the classical limit, a consistent analysis
must take into account the constraints of quantum mechanics, in
particular the uncertainty principle. That, and the fact that energy is a gauge charge,
is all  will use in this paper.

Purely for the purposes of being able to visualize the geometry, I
am going to analyze the question in 2+1 dimensions.
Let us begin by describing a two dimensional  spatial geometry
containing a wormhole. We take the usual large sheet of paper and
bend it so that distant points are close in three dimensional
embedding space. Then we cut two mouths and identify the edges. If
the long way around is very big we can ignore the fact that the
two sheets are connected (except by the wormhole) and think of
them as two separate universes connected by a wormhole. Call the
two sheets $\bf A$ and $\bf B$. We live on $\bf A$.

Seen from either side, the wormholes may have mass. If so they create a gravitational
field which in 2+1 dimensions means a conical deficit. In that
case the geometry would be a pair of cones connected at their tips
by a smoothed region of negative curvature. Incidently, the fact
that the wormhole is negatively curved is an explicit
demonstration that negative energy is needed to support the
configuration.

 \setcounter{equation}{0}
\section{The Electric Case}

Let's begin with a simple experiment involving two boxes (boxes $\bf 1,2$) that can contain electrically
charged particles. Call the particles $\pi^+$ mesons. We can construct a state (we work in Coulomb gauge)
with total charge $N$ and relative phase $\theta$.
\be
|N \theta \rangle = \sum_{n=0}^N |N-n, n\rangle e^{in\theta }
\ee
where $|m, n\rangle$ means a state with charge $m$, and $n$ in box $\bf 1$ and $\bf 2$.
The relative phase is a measurable \cite{aharonov} and can be measured as follows. A beam of neutrons
is split into two beams. One beam is sent through box $\bf 1$ and the other through box $\bf 2$. In passing through
the box of $\pi^+$ mesons, a charge may be absorbed, turning a neutron into a proton. When the beams are
recombined, the probability to find a proton contains an interference term
proportional to $\cos (2 \theta)$.

Typically the phase will not be time
independent but, because of the electrostatic coulomb energy, will oscillate in a regular predictable way.
The boxes can be separated with the phase relations remaining intact.

Imagine the two boxes  starting out on our side of the wormhole on sheet $\bf A$. Now transport box
$\bf 2$ to the other end of the wormhole, taking it the long way around. At the end of this process
we have two charge reservoirs, one on each sheet, with definite relative phase.
this provides us with a means of
comparing phases at the two mouths.

Next, pass box $\bf 1$ through the wormhole so that it appears on sheet $\bf B$. Now both boxes
are on the same sheet and presumably have a definite phase that can be compared.
But in fact when we experimentally compare the phases we find them to be completely
uncorrelated: the interference term proportional to $\cos (2\theta)$ is absent.

To see why this is so, lets examine more closely what happens when a charge is passed
through a wormhole. When the charge began on sheet $\bf A$, its electric flux lines radiated out
to infinity on that side. As we pass the charge through the wormhole, its flux lines do not
suddenly asymptotically rearrange. They thread back through the wormhole and continue to radiate
to infinity on sheet $\bf A$. In other words NO CHARGE passed through the wormhole.
In fact what the observer on side $\bf B$ sees is that in addition to the charge that passed through,
an opposite charge has formed on the wormhole mouth. On side $\bf A$ the charge is also unchanged but
 it now exists in the form of a charged wormhole mouth.

 Now lets return to the problem of two charged boxes with definite relative phases. After sending box
 $\bf 1$ through the wormhole, there are four charged systems to keep track of--the two boxes and the two
 wormhole mouths, $\bf A$ and $\bf B$. Instead of finding the phases of the two boxes correlated, what one finds
 is a phase correlation between box $\bf 2$ and mouth $\bf A$, and
 a similar correlation between the phases of $\bf 1$ and $\bf B$, but no correlation between phases $\bf 1$ and $\bf 2$.

 \setcounter{equation}{0}
\section{Time and Energy Transfer}

Energy and time are related in the same way as charge and phase. Let's
repeat the analysis and see what it says for this case. For
ease of visualizing the system, we work in 2+1 dimensions but
the analysis applies in higher dimensions.

First consider what happens when a mass passes through the wormhole from $\bf A$ to $\bf B$.
For simplicity consider a point
mass. The analogue of the lines of electric flux is the conical deficit that extends to infinity
on sheet $\bf A$. As in the electric case, when the mass passes through the wormhole, the conical
deficit remains behind and no change takes place in the deficit on either side.
On side $\bf A$ the deficit appears to be due to mouth $\bf A$. On side $\bf B$ the deficit is shared between
the mass point and mouth $\bf B$, which appears to have lost whatever mass is carried by the mass
point that it spit out.

Let's start over. Instead of two boxes of charge that serve as fiducial phases, we take a given energy
and partition it into two clocks, call them $\bf 1$ and $\bf 2$. The phase variables are
replaced by the time readings $t_1, t_2$.   The clocks can be initially prepared so that the time
difference, $t_1- t_2$ is zero: in other words the clocks are synchronized.

Next transport clock $\bf 2$ the long way to sheet $\bf B$. If this is done with no
great velocity and without passing through gravitational potential wells, the two clocks will
remain synchronized. In any case there is no difficulty in accounting for any difference in proper
times along the respective paths of the clocks.
We now  have clocks that can compared near the two mouths.

Here is the experiment that we would like to do. Pass
 clock $\bf 1$ through the wormhole and compare it with clock $\bf 2$. After recording the result,
clock $\bf 1$ may return through the wormhole and report its finding. One would naively expect that apart from effects that might occur in the wormhole,
the two clocks should agree. In other words the relative time observable $(t_B-t_A)$ would be
expected to be very small compared to $L$.
But it is clear that this cannot be so. Instead the two clocks will be completely uncorrelated.
This is a consequence
of the energy-time uncertainty principle and the fact that the energy
that passed through the wormhole was exactly zero. Thus we can conclude that
\be
\langle(t_B-t_A)^2\rangle =\infty.
\ee
Note that clock $\bf 1$ is equally likely to come out in the infinitely remote past as in
the infinitely remote future!

Another way to say what happened is that when  clock $\bf 1$ entered and exited at $\bf A$ and   $\bf B$,
it left behind clocks at the wormhole mouths. After exiting at $\bf B$, clock $\bf 1$ found itself synchronized
with the  clock at mouth $\bf B$, but totally random with respect to clock $\bf 2$.

\setcounter{equation}{0}
\section{Conclusion}

Note added:

The conclusion described below has been seriously criticized in a recent rebuttal.  See gr-qc/0504039.

Traversable wormholes that would allow an observer to short-circuit large space-like distances by
entering one mouth and exiting another violate quantum mechanics. The average magnitude of the
time between
entering and exiting is infinite as a consequence of the uncertainty principle and the fact that
the total ADM energy transfer must be exactly zero.

The only sensible interpretation is that the events at the two wormhole mouths are completely
uncorrelated. The appearance of a clock similar to clock $\bf 1$  on sheet $\bf B$ can be interpreted as
a quantum emission by an unstable object, namely the mouth $\bf B$. That decay can take place
at any time and is uncorrelated to
the ``absorption" event at $\bf A$.  I think it is even justified to say that
the observer never left $\bf A$.
It was left behind in the form of the complex energy levels of  mouth $\bf A$.

As for time travellers, they will need to find another kind of time machine.

\setcounter{equation}{0}
\section{Note added}
Serguei Krasnikov has reminded me that there is an additional
parameter needed to specify a wormhole, namely a relative time variable
that specifies how the identification is done. For example two stationary
wormholes mat be identified at equal time in the rest frame or there may
be an offset by some finite delay or advance. Obviously if the delay is finite
this will make no difference to the argument, even if the delay time is somewhat
uncertain, quantum mechanically.

One way of fixing the time delay is to start with a wormhole that is short both the
"long way" and through the wormhole. Some authors have speculated that such small
wormholes might be created in the lab. Once created, they can be slowly separated to a
large distance. If separated symmetrically, the delay time must be zero by symmetry.
Moreover, there is no reason why the delay would fluctuate beyond whatever small
fluctuation may have been there in the beginning.

\end{document}